\documentclass[aps,prd,longbibliography,reprint,superscriptaddress,super,amsmath,showkeys,notitlepage]{revtex4-1}

\usepackage[colorlinks,linkcolor=blue,bookmarks=true,citecolor=blue,urlcolor=blue,breaklinks=true]{hyperref}
\usepackage[T1]{fontenc}
\usepackage[utf8]{inputenc}
\usepackage[]{graphicx}
\usepackage{amssymb,enumitem,physics,cleveref,soul}
\usepackage[table,usenames,dvipsnames]{xcolor}
\usepackage{comment}
\newcommand{\nn}{\nonumber}

\newcommand{\Oop}{\hat{\rho}}
\newcommand{\Hop}{\hat{H}}
\newcommand{\Sop}{\hat{S}^{-}}
\newcommand{\Sopz}{\hat{S}_z}
\newcommand{\Sdop}{\hat{S}^{+}}
\newcommand{\aop}{\hat{a}}
\newcommand{\adop}{\hat{a}^\dagger}
\newcommand{\sigop}[2]{\hat{\sigma}^{#1}_{#2}}
\newcommand{\lk}{\left(}
\newcommand{\rk}{\right)}
\newcommand{\lsz}{\left[}
\newcommand{\rsz}{\right]}
\newcommand{\lka}{\left\{}
\newcommand{\rka}{\right\}}

\begin{document}
\title{Superquantization rule for multistability in driven-dissipative quantum systems}

\date{\today}
 \author{Nikolett Német}
 \email{nemet.nikolett@wigner.hun-ren.hu}
 \affiliation{HUN-REN Wigner RCP, H-1525 Budapest, P.O. Box 49., Hungary}
 \author{Árpád Kurkó}
 \email{curko.arpad@wigner.hun-ren.hu}
 \affiliation{HUN-REN Wigner RCP, H-1525 Budapest, P.O. Box 49., Hungary}
 \author{András Vukics}
 \email{vukics.andras@wigner.hun-ren.hu}
 \affiliation{HUN-REN Wigner RCP, H-1525 Budapest, P.O. Box 49., Hungary}
 \author{Péter Domokos}
 \email{domokos.peter@wigner.hun-ren.hu}
 \affiliation{HUN-REN Wigner RCP, H-1525 Budapest, P.O. Box 49., Hungary}

%\email[Correspondence and requests for materials should be addressed to, e-mail address: ]{}

\begin{abstract}
We present a superquantization rule which indicates the possible robust stationary states of a generic driven-dissipative quantum system. Multistability in a driven cavity mode interacting with a qudit is revealed in this way within a simple intuitive picture. The accuracy of the superquantization approach is confirmed by numerical simulations of the underlying quantum model. In the case when the qudit is composed of several two-level emitters coupled homogeneously to the cavity, we demonstrate the robustness of the superquantized steady states to single-emitter decay.
\end{abstract}

%\setboolean{displaycopyright}{true}
\pacs{<pacs codes>}
\maketitle

Out-of-equilibrium quantum phases can exist in a generic class of microscopic systems formed by a single driven-dissipative  harmonic oscillator coupled to a small, finite dimensional quantum system \cite{abraham1982optical,lugiato1984ii,reinisch1994optical,Kilin1993MultistabTavis,Carmichael2015PBB,Gutierrez2018GeneralJCRabi,Hwang2018,Zhang2021,reiter2020cooperative}. The external coherent drive balanced by the loss to the environment leads to a steady state with a stationary current of quanta through the system that provides for a measurable signal \cite{Fink2017ObsPBBT,Suarez2023,gabor2023ground,sett2024emergent}. The components of the steady state are unambiguously identified through the current outcoupled from the oscillator mode by a classical measurement apparatus. The defining feature of these quantum phases is that they show resilience to the back-action of measurement \cite{berdou2023one}. All this is possible because one of the two components is a harmonic oscillator which, when dressed by the coupling to the small system, forms highly excited stationary quasi-coherent states \cite{Alsing1991,kurko2023photon}.

A typical implementation can be an electromagnetic radiation mode in a resonator, being the oscillator, while a set of (artificial) atoms coupled to this mode constitutes a non-linear quantum medium \cite{brune1996quantum,fink2008climbing,Bishop2009,kasprzak2010up}. The measurement signal is provided by the field outcoupled from the resonator which is recorded by a photo-detector. While the harmonic oscillator ensures the robustness of the stationary states, the number and characteristics of the possible out-of-equilibrium phases are determined by the properties of the small finite-dimensional system strongly coupled to the oscillator \cite{Alsing1991,Shamailov2010}, as is shown in this paper through a simple example.

The simplest possible scenario is when a two-level system is coupled to a coherently driven harmonic oscillator. The quantum mechanical description of such a setup is given by the driven Jaynes-Cummings model \cite{Larson2022}. This system is readily realized in cavity-QED settings, i.e., as a qubit coupled to a single mode of a resonator. Previously, two components of the steady-state have been identified in such a system, that can exists in a mixture in a “bistability” domain of the parameter space \cite{Savage1988,Carmichael2015PBB, Dombi2015BistabXtrStrong, Gutierrez2018GeneralJCRabi, Fink2018, Curtis2021}. One of them is referred to as the “dim” state, where the photon transmission through the resonator is suppressed due to the lack of resonance with the uneven energy-level spacings of the low-energy part of the spectrum. The other one is referred to as the “bright” state where the cavity-QED system is highly transmissive, emitting a strong photo-current from the driven resonator.

The two-component steady state manifests on quantum trajectories as stochastic alternation between the two metastable states, resulting in a telegraph signal in the resonator intensity \cite{Vukics2019FiniteScalePBBT}. The mechanism of the transition from the dim to the bright state has been termed the breakdown of photon blockade \cite{Carmichael2015PBB,Dombi2015BistabXtrStrong,Vukics2019FiniteScalePBBT,Curtis2021,chiorescu2010magnetic,Palyi2012}. It has also been shown that the switching behavior is a finite-size precursor of a first-order dissipative phase transition \cite{Carmichael2015PBB}. Increasing the strength of the coupling between the qubit and the mode leads toward a thermodynamic limit where the dwell times of the metastable states and their separation in mode intensity become infinite \cite{Carmichael2015PBB,Vukics2019FiniteScalePBBT,Curtis2021,sett2024emergent}. The macroscopic switching of output intensities has been experimentally demonstrated in circuit-QED with transmon artificial atoms \cite{Fink2017ObsPBBT,sett2024emergent}. Transmons are intrinsically multilevel structures; therefore, realistic characterization potentially requires multiple energy levels, and circuit QED systems are non-linear par excellence \cite{Rebic2009,peano2010dynamical,Dykman2012,fitzpatrick2017observation,mavrogordatos2017simultaneous,Muppalla2018,heugel2019quantum,Wustmann2019,brookes2021critical}.

\begin{figure*}
\centering
    \includegraphics[width=1\linewidth]{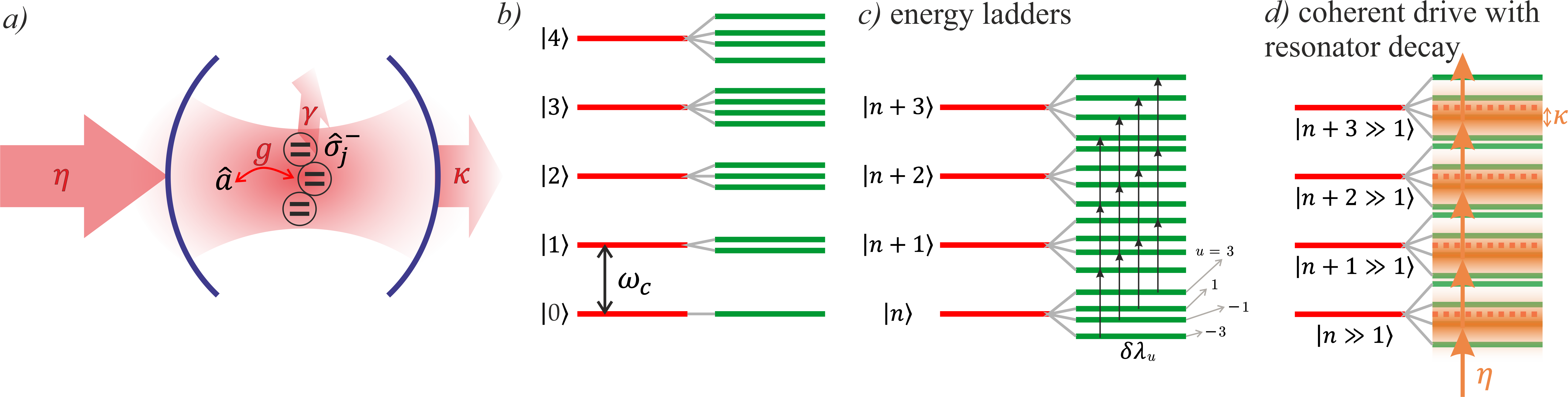}
\caption{Left: a spin-3/2 is composed of three qubits coupled with the same strength $g$ to a driven resonator mode with linewidth $\kappa$. a) In the spectrum of the Hamiltonian without the drive ($\eta=0$ in \cref{eq:Ham}), the first two excited states split into a doublet and a triplet, respectively. b) Higher up on the ladder they split into quadruplets. c) When driving is considered, the detuning of the drive from the mode and spin transition frequencies determines which eigenstates are involved in shaping the resonator output intensity.}
\label{fig:energies}
\end{figure*}

In this paper, we consider a spin-3/2 coupled to a single-mode resonator field (cf. \cref{fig:energies}) in order to reveal the connection between the structure of the non-linear atomic system and the possible multistable solutions \cite{landa2020multistability}. We will show that these solutions can be determined from a simple algebraic self-consistency equation, i.e., a “super-quantization” rule. This equation is derived on the basis of an intuitive picture of the emergent quasi-coherent states and proves to be very accurate, as it is also demonstrated using quantum-trajectory simulations of the full quantum dynamics of the driven-dissipative system using the C++QED framework \cite{sandner2014cpp,*cppqedgithub}. In the case when the spin-3/2 is composed of three qubits coupled homogeneously to the mode, we can add single-qubit decay to the quantum trajectories, and demonstrate the robustness of the quasi-coherent states to this decoherence channel. This is a remarkable result, since the quasi-coherent states, being derived solely from the Hamiltonian, reside in the completely symmetric subspace, which is not closed against single-emitter decay.

%\section{Intuitive analytic description}
The system Hamiltonian is the driven Tavis-Cummings model \cite{Tavis1968,Tavis1969,fink2009dressed} that in a frame rotating with the drive frequency reads
\begin{equation}
\label{eq:Ham}
\Hop=-\Delta\lk\adop\aop+\frac\Sopz2\rk+
ig\lk \adop\Sop-\Sdop\aop\rk+i\eta\lk\adop-\aop\rk,    
\end{equation}
where $\aop$ and $\adop$ are the mode annihilation and creation operators, the $\hat{S}$s are the spin operators, and $g$ is the spin-resonator coupling strength. The resonator mode is tuned to the spin transition frequency. Finally, $\eta$ is the amplitude of the coherent resonator driving, with detuning $\Delta$ from both the frequencies of the spin and the mode. We consider photon loss from the resonator described by the Liouvillian operator
\begin{equation}
\label{eq:dissip}
    \mathcal{L}\lsz\Oop\rsz=\kappa\lk 2\aop\Oop\adop-\adop\aop\Oop-\Oop\adop\aop\rk%+\nn\\
%    &\gamma\sum_{j=1}^N\lk2\sigop{-}{j}\Oop\sigop{+}{j}-\sigop{+}%{j}\sigop{-}{j}\Oop-\Oop\sigop{+}{j}\sigop{-}{j}\rk,
\end{equation}
where $2\kappa$ is the rate of resonator decay% and $2\gamma$ is the spontaneous emission rate of the qubits (atoms) to the surrounding  modes other than that of the resonator
. The loss of excitations are compensated by the coherent field that drives the system into a steady state.

In order to obtain an intuitive picture of the dynamics, we first consider only the undriven Hamiltonian, that is, we take $\eta=0$ in \cref{eq:Ham}. The eigenstates of this system have been studied in detail \cite{Shen2014,Zhang2016}. Since the Hamiltonian conserves the total excitation number $n$, it is enough to perform the diagonalization on a subspace spanned by $\ket{n-\lk m+\tfrac32\rk}\ket m$, where $m\in\lka -\tfrac{3}{2},-\tfrac{1}{2},\tfrac{1}{2},\tfrac{3}{2}\rka$ is the spin-projection quantum number. The eigenvalues read (cf. \cref{fig:energies}[a-b])
\begin{subequations}
\begin{align}
\label{eq:lambdas}
    \lambda_{-3}(n)=&-\Delta\,n-g\sqrt{(5+4\sqrt{1+\epsilon^2})(n-1)},\\
    \lambda_{-1}(n)=&-\Delta\,n-g\sqrt{(5-4\sqrt{1+\epsilon^2})(n-1)},\\
    \lambda_{1}(n)=&-\Delta\,n+g\sqrt{(5-4\sqrt{1+\epsilon^2})(n-1)},\\
    \lambda_{3}(n)=&-\Delta\,n+g\sqrt{(5+4\sqrt{1+\epsilon^2})(n-1)},
\end{align}
\end{subequations}
where $\epsilon=\frac{3}{4(n-1)}$ vanishes for large enough photon numbers, so that even for $n\sim 3$, $\epsilon^2 \ll 1$, and the eigenvalues can be well approximated by 
\begin{equation}
\lambda_{u}(n)\approx-\Delta\,n+u\,g\sqrt{n},
\label{eq:m_j}
\end{equation}
with dressed-state ladder indices $u\in\lka -3,-1,1,3\rka$. The lower part of the spectrum is “qubit-dominated”,  meaning that it is highly non-linear. For high photon numbers $n$, the frequency difference between the dressed states corresponding to given $u$ in adjacent manifolds  $n$ and $n+1$ is
\begin{align}
\label{eq:dellamb}
    \lambda_u(n+1)-\lambda_u(n) &\approx -\Delta+\frac{u\,g}{2\sqrt{n}}.
\end{align}
The function $1/\sqrt{n}$ is slowly varying such that there is a range in which the difference is almost constant, i.e., the dressed states with identical $u$ in the high $n$ manifolds span a closely equidistant ladder. This part of the spectrum hosts quasi-coherent states that can be steadily maintained by the external drive $\eta$.

This is the ground for forming an intuitive picture. The equidistant ladders at high numbers $n$ can be considered as harmonic oscillators. The well-known solution of an externally driven, lossy harmonic oscillator leads then to the self-consistent equation for the mean photon number, 
\begin{equation}
\label{eq:n}
n=\frac{\eta^2}{\kappa^2+\lsz\lambda_u(n+1)-\lambda_u(n)\rsz^2}= \frac{\eta^2/\kappa^2}{1+\lk\delta-\frac{u\,g}{2\kappa\sqrt{n}}\rk^2},
\end{equation}
with $\delta=\Delta/\kappa$. Since \cref{eq:n} leads to a quadratic polynomial equation in $\sqrt{n}$, at a given detuning, there can exist two distinct solutions $n$ for each ladder $u$. However, it can be shown that the secondary root is non-physical, since it decreases with increasing drive amplitude \footnote{A monotonically decreasing intra-cavity intensity vs. drive amplitude function means that if for any finite drive amplitude the intensity is positive, then the intensity remains non-zero for zero drive, which is manifestly non-physical. On the other hand, negative intensity (for any drive strength) is also clearly non physical.}.

The possible maximum number of stationary bright states is thus 4 in the case when the qudit interacting with the mode is a spin-3/2. From the way this result has been derived above, it is clear that this number equals the number of dimensions of the qudit. For a given drive amplitude, however, not all of these solutions are realized over the full quantum dynamics of the driven-dissipative system, since the solution has to fulfill other conditions that we detail below. The maximum intra-cavity photon number is  $\eta^2/\kappa^2$, reached at exact resonance, $\delta= {u\,g}/{2\eta}$. In this case, the qudit tunes the cavity back into resonance with the driving field. 

From the same intuitive picture of the driven-dissipative steady state of what is approximately a harmonic oscillator, we can obtain a self-consistent equation for the complex intra-cavity field amplitude, that is more general than \cref{eq:n}, as it gives information also on the phase \cite{Alsing1991,armen2009spontaneous}:
\begin{equation}
\label{eq:alpha_m}
\alpha_u =\frac{\eta/\kappa}{1-i\lk\delta-\frac{u\,g}{2\kappa|\alpha_u|}\rk}.
\end{equation}
%The same description was applied in \cite{Fink2017ObsPBBT,Fink2018,sett2024emergent} for up to three qubits in the resonator. 
This is a kind of quantization rule, i.e., a “super-quantization” for the quasi-coherent field amplitudes. Straightforward calculation for the imaginary and real parts,
\begin{equation*}
\alpha_u \equiv |\alpha_u|\lk\cos{\varphi_u}+i\sin{\varphi_u}\rk,
\end{equation*}
leads to the following expressions
\begin{subequations}
 \label{eq:intuitive} 
\begin{align} 
    |\alpha_u|&= \frac{\eta}{\kappa}\cos{\varphi_u} \label{eq:intuitivea}\\
    \cos{\varphi_u}&=\frac{1}{\sqrt{1+\delta^2}}\lk\zeta_u\delta + \sqrt{1-\zeta_u^2}\rk
    \label{eq:intuitiveb}\\
    \sin{\varphi_u}&=\frac{1}{\sqrt{1+\delta^2}}\lk-\zeta_u + \delta\sqrt{1-\zeta_u^2}\rk
    \label{eq:intuitivec}
\end{align}
\end{subequations}
where we introduced the scaled drive strength parameter
\begin{align}
\label{eq:zeta}
    \zeta_u=\frac{u\,g/2\eta}{\sqrt{1+\delta^2}} \, .
\end{align}

\begin{figure}
\centering
\includegraphics[width=0.99\linewidth]{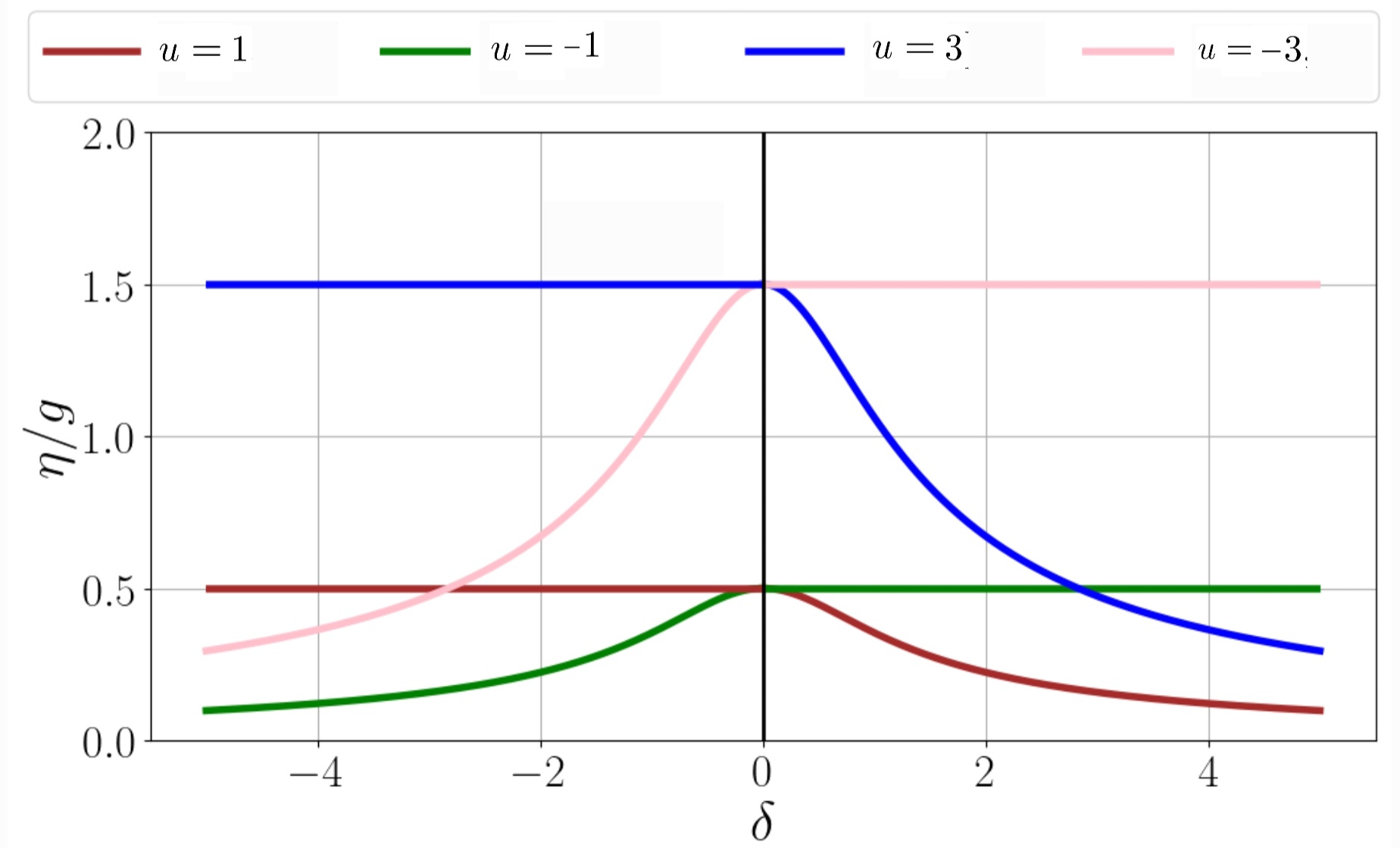}
\caption{Map of the available solutions according to the intuitive approach, illustrated in the plane of the external control parameters of the drive $\eta$ and $\delta$. The lines represent the boundaries above which the solutions with the applied colours become physical.}
\label{fig:zeta_map}
\end{figure}
In contrast to the implicit expression for the mean photon number in \cref{eq:n}, here we got an explicit result for the cavity field amplitude determined by the external control parameters, i.e., by the drive amplitude $\eta$ and the detuning $\delta$. Such a result may correspond to a valid physical steady state provided the condition $0\leq\cos{\varphi_u}\leq 1$ is satisfied. This is summarized in the solution diagram of \cref{fig:zeta_map}. Clearly, the drive strength parameter $\zeta_u$ must stay below a threshold to ensure that the solution \labelcref{eq:intuitivea} is both real and positive.
\begin{itemize}
    \item In the case of $\zeta_u\delta \ge 0$ this is $\abs{\zeta_u} \le 1 $
    \item and for $\zeta_u\delta < 0$, $\abs{\zeta_u}\sqrt{1+\delta^2} \le 1 $.
\end{itemize}
In cases where these conditions are not satisfied, the solution for the intra-cavity photon number corresponding to $u$ does not exist. Furthermore, even if the solution exists according to \cref{eq:intuitive}, it is not necessarily consistent with the assumption of large enough photon numbers and, accordingly, a quasi-equidistant ladder $u$.

Let us now turn to the question how relevant the intuitive solutions are for the fully quantum driven-dissipative dynamics. We are generating quantum trajectories based on the quantum-jump Monte Carlo method implemented in the C++QED framework.

\begin{figure*}
\centering
\includegraphics[width=0.95\linewidth]{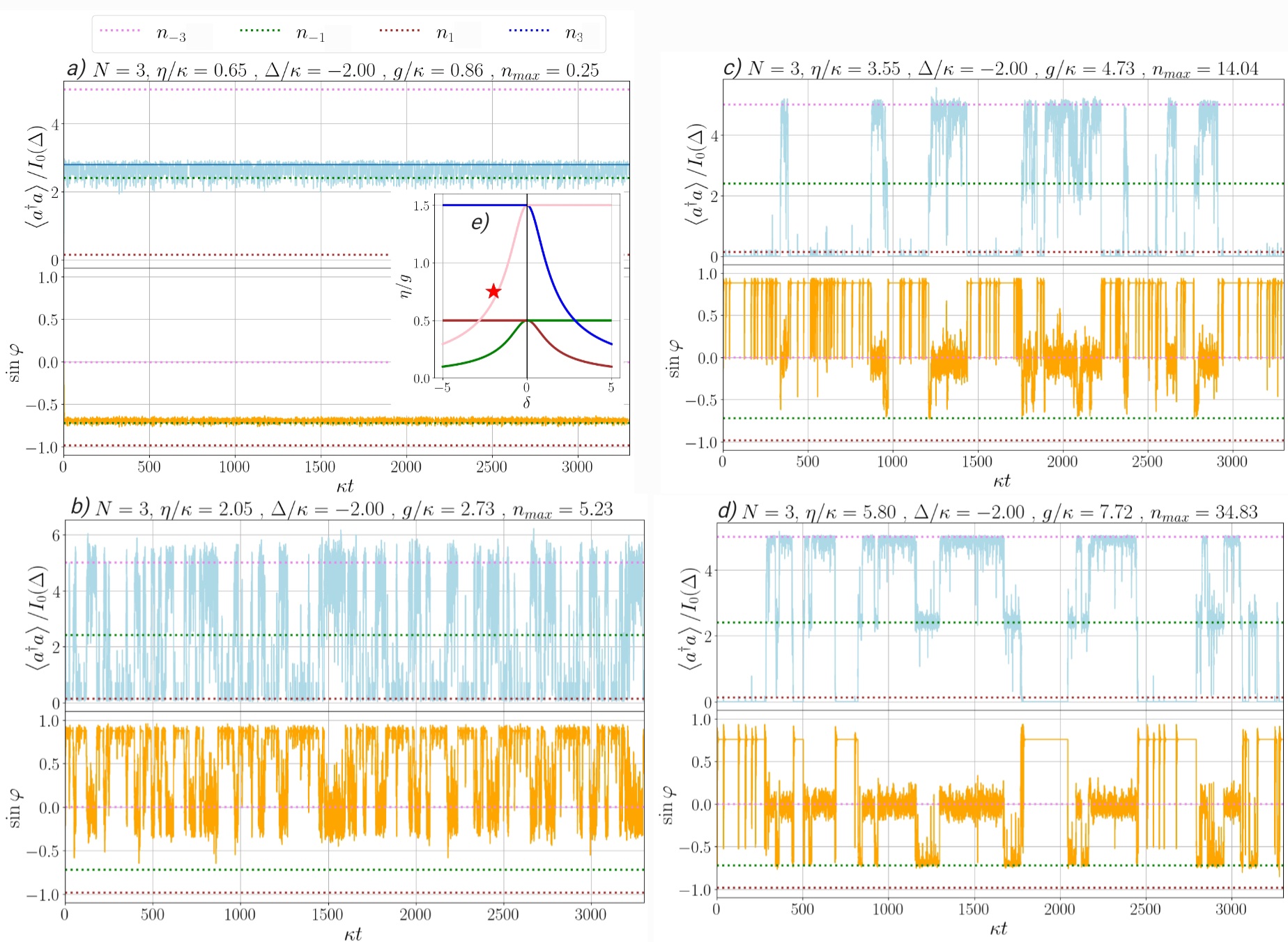}
\caption{Quantum trajectories for the cavity field amplitude. Numerically calculated time evolution of the cavity photon number normalized to the empty cavity photon number (blue, upper half of the panels), and phase (orange, lower panels). The fixed values $\zeta_{u}/u =0.6$ and $\delta=-2$ determine the intuitive solutions in \cref{eq:intuitive}, indicated by the horizontal dotted lines with color codes. The subplots (a)-(d) correspond to different coupling strength $g$ and drive amplitudes $\eta$ such that  $g/\eta$ is constant. The four panels display examples for the photon transmission in the weak (a), in the intermediate (b), and in the strong coupling regime with bistability (c) and multistability (d). The inset (e) shows the chosen work point on the map of possible quasi-coherent solutions, cf. \cref{fig:zeta_map}.}
\label{fig:quantumtrajectories}
\end{figure*} 

Sample quantum trajectories are presented in Figure \ref{fig:quantumtrajectories} for fixed values of parameters $\zeta_u/u=0.6$ and $\delta=-2$ which determine the intuitive solutions, indicated by dotted horizontal lines in the panels. All the solutions except for $u=3$ are possible in this case, however, the solution for $u=1$ is at very low photon numbers, contradicting the premise of the intuitive approach. The different panels are devoted to illustrate the emergence of the predicted quasi-coherent solutions as the coupling strength increases, together with the drive strength $\eta$ such that $g/\eta$ is constant. With this scaling the intuitive solutions are fixed relative to the empty-cavity intensity, but the higher drive strength can inject higher photon numbers in the cavity. Accordingly, the quantum trajectories exhibit substantially different behaviour regarding the cavity transmission for the four selected values of the coupling constant $g$ in the panels (a)-(d).

For the case of weak coupling, shown in panel (a), the low-lying excitation spectrum does not lead to transmission blockade \cite{imamoglu1997strongly,lang2011observation} at all due to weak anharmonicity. The qudit, in its low excitation regime being a linear polarizability,  amounts to a simple dispersive shift of the mode resonance that results in an approximately steady intra-cavity photon number $\langle \adop \aop \rangle / I_0(\Delta) = 2.8$. This average over the numerical trajectory is fit well by the value calculated from the dispersive shift, represented as the blue solid horizontal line in panel (a), though this shift is estimated by a leading order perturbative calculation. Throughout, we use for normalization the empty-cavity intensity $I_0(\Delta)=\frac{\eta^2}{\kappa^2+\Delta^2}$. None of the potential quasi-coherent solutions for the fixed value of $\zeta_u/u = 0.6$ play any role in this parameter regime. The “super-quantization rule”  \labelcref{eq:alpha_m} is applicable only at high photon numbers which is not met in panel (a). The range of photon numbers actually encountered in the time evolution is indicated by the maximum photon number $n_\text{max}$ in the row of parameters above the panels, i.e., $n_\text{max}=0.25$ for (a). The photon number is not high enough in panel (b), either, which represents an intermediate range of the coupling, $g = 2.73 \kappa$. There is still no well-resolved photon blockade, and the intuitive solutions do not appear, either. The system does not find a robust attractor state and thus strong fluctuations of the photon number fill the full range of intensity between vacuum level and the $n_\text{max}=5.23$.

Going toward the strong coupling regime \cite{forn2019ultrastrong}, in panel (c), the intuitive solution with $u=-3$ emerges clearly as a metastable solution for macroscopic durations of time, alternating with photon blockade periods in a random switching signal. This is a similar telegraph signal of bistability as the one studied with a single qubit \cite{Vukics2019FiniteScalePBBT}. The quasi-coherent field state according to the super-quantization rule is formed here at photon numbers around $\sim 14$. With still stronger coupling, in panel (d), multiple solutions can stabilize and can be observed in the intra-cavity intensity. The appearance of these metastable solutions with the increasing parameter $g$ confirms that the strong-coupling “thermodynamic limit” \cite{Carmichael2015PBB} holds for the breakdown of photon blockade in a qudit-mode system as well. The numerical study furthermore verifies that the phase of cavity field, when the system state stabilizes in one of the available quasi-coherent states, follows accurately the analytical value in \cref{eq:intuitive}.

\begin{figure}
\centering
\includegraphics[width=0.87\linewidth]{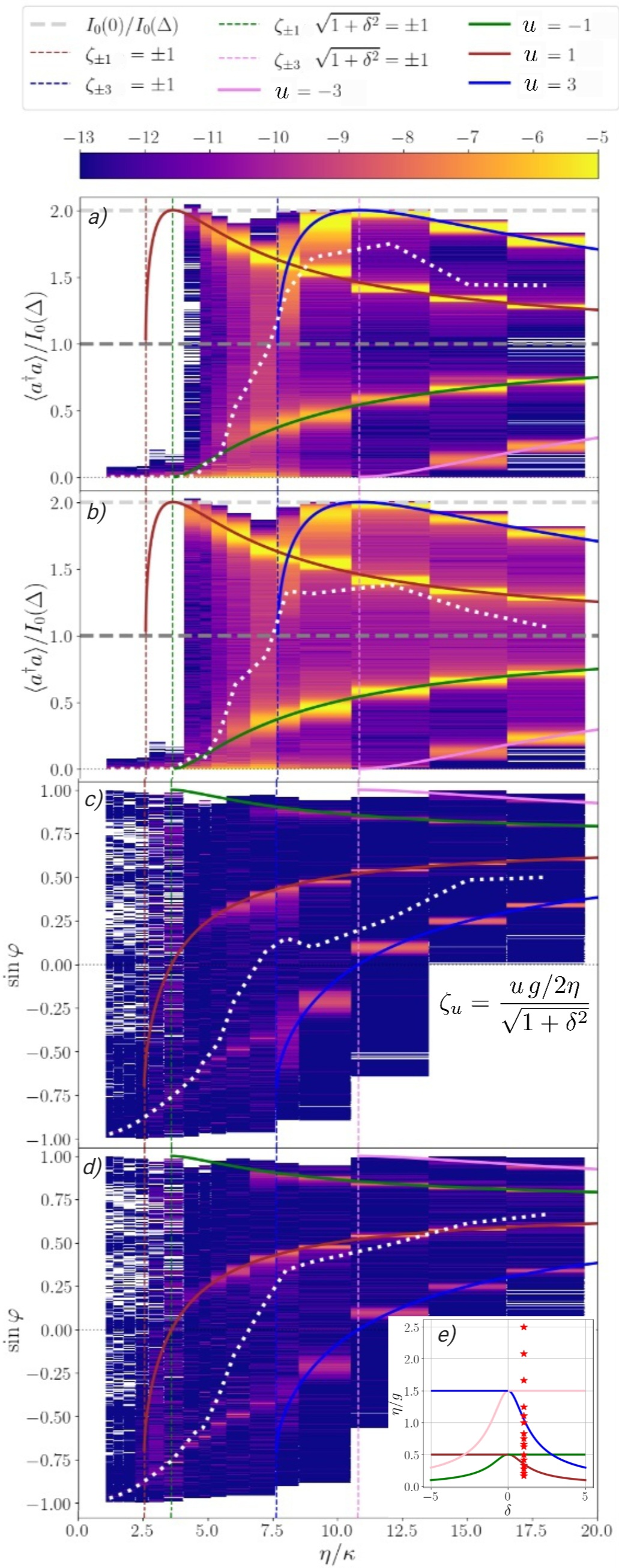}
\caption{Histograms showing the probabilities of normalized resonator intensity (with respect to the detuned empty-cavity intensity) (a-b) and phase values (c-d). The parameters $g=7.2\kappa$ and $\Delta=\kappa$ are fixed, while the drive strength $\eta$ is varied, corresponding to points on the right side of (e). Qubit spontaneous emission is considered in (b) and (d) with rate $\gamma=\kappa/100$. The field observables gather around the solutions of \cref{eq:intuitive} which gradually appear as solid lines, the critical drive strengths associated with the emergence of different $u$ solutions are depicted as vertical dashed lines. The white dotted lines connect the time averages of the corresponding observable at the given drive intensities.}
\label{fig:hists}
\end{figure}

The demonstration of the adequacy of the intuitive picture in the strong-coupling regime is completed in \cref{fig:hists}. Here, the coupling constant $g$ is fixed at a given detuning $\delta=1$, and the drive strength $\eta$ is increased in order to address increasingly higher lying dressed states. Long quantum trajectories have been generated, and the extracted photon-number histograms are displayed in color code. 

\begin{figure}
\centering
\includegraphics[width=0.95\linewidth]{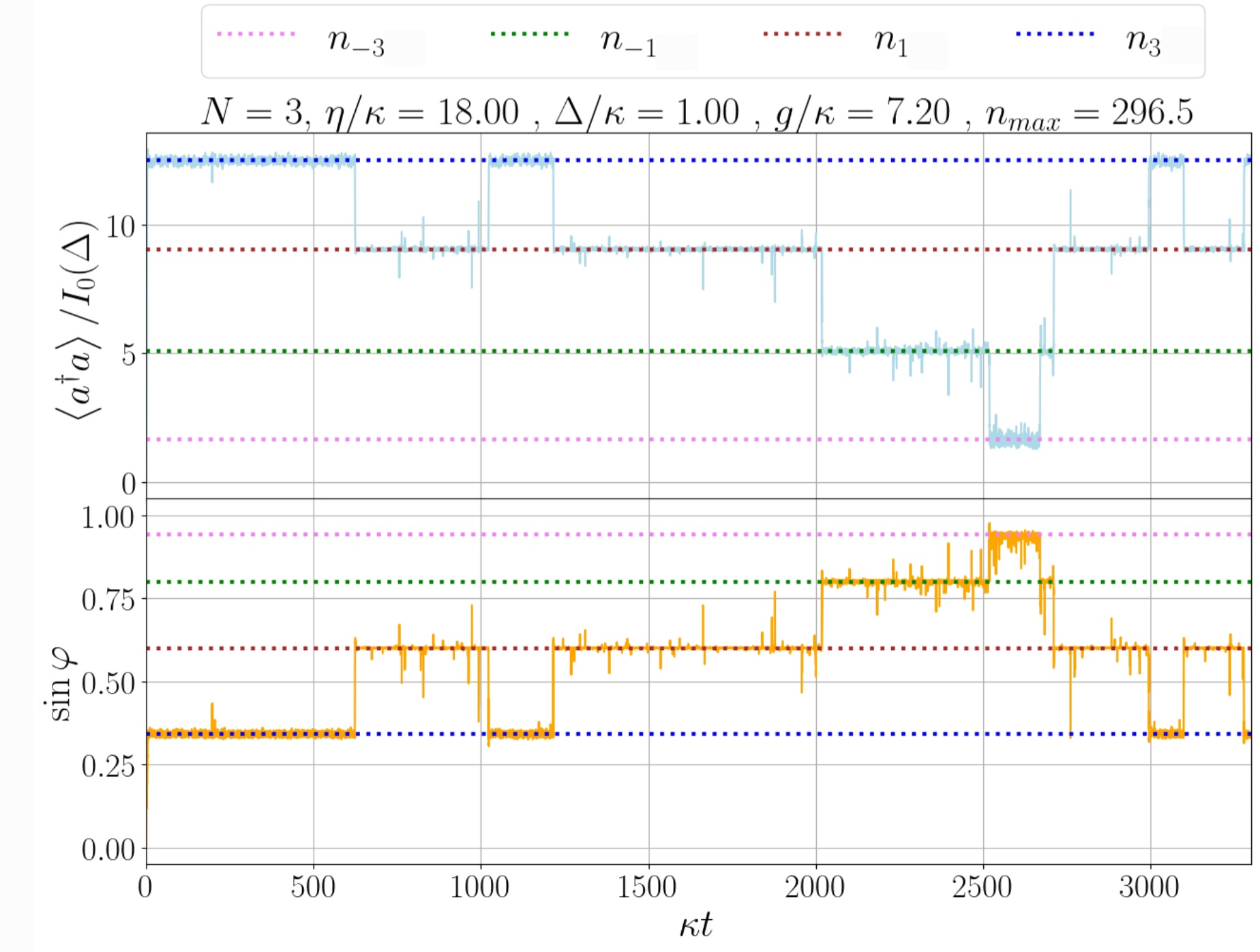}
\vglue -.2cm
\caption{Example trajectory where all the solutions are present. Parameters: $\gamma=0, g=7.2\kappa, \eta=18\kappa$ and $\Delta=\kappa$ }
\label{fig:exxtra}
\end{figure}

For weak drive the cavity is in a dim state due to the drive being out of resonance with the dressed excitation levels in the strong coupling regime. This is the photon blockade regime which is broken down at a certain drive strength. A new peak at around $\langle \adop \aop \rangle / I_0(\Delta) = 2$ appears in the histogram, cf.~\cref{fig:hists}(a). Vertical dashed delimit the onset of a new intuitive solution associated with a ladder $u$,  where this critical value of $\eta$ is determined by the condition $\zeta_u \le 1$ ($\abs{\zeta_u}\sqrt{1+\delta^2} \le 1$) for $u>0$ ($u<0$). On increasing the drive intensity, the numerical simulation confirms the emergence of further new peaks in the photon number distribution. For very strong  drive, i.e. the right side of the panels, all the four possible solutions predicted by the intuitive picture for $\delta>0$, cf.~\cref{fig:zeta_map}, are occupied in a genuine multistability. A sample quantum trajectory with cavity intensity switching among the four possible values is shown in \cref{fig:exxtra}. The position of the peaks in the photon number cf.~\cref{fig:hists}(a) as well as in the phase distribution cf.~\cref{fig:hists}(c) fit very accurately on the intuitive solution represented by solid lines with various colours corresponding to the different ladders $u$.  

The stability of the quasi-coherent states is due to their robustness against the cavity photon loss, just as a true coherent state of an oscillator is robust against a photon-loss quantum jump. In our case, a quantum jump corresponding to the loss of a photon mixes negligibly the ladders corresponding to different $u$s. This is why the drive (coherent displacement) and the dissipation establish a stable Poissonian distribution of excitations in one of the ladders.

On panels (b) and (d) of \cref{fig:hists}, a new piece of physics is added. For these plots, we considered the spin-3/2 as being composed of 3 qubits with identical transition frequencies, and homogeneously coupled to the mode, that is $\Sop=\sum_{i=1}^3\sigop-i$ in \cref{eq:Ham}. In this case, it becomes a relevant question what the effect of single-qubit decay is on the dynamics, that we can simulate by supplementing the dissipator \labelcref{eq:dissip} with
\begin{equation}
\label{eq:dissipse}
    \mathcal{L}_\text{single emitter}\lsz\Oop\rsz=\gamma\sum_{i=1}^3\lk 2\sigop-i\Oop\sigop+i-\sigop+i\sigop-i\Oop-\Oop\sigop+i\sigop-i\rk.
\end{equation}
What is remarkable is that this dissipator brings out of the subspace determined by the spin quantum number $3/2$, that is, where the state of the qubits is permutationally symmetric. Single-emitter decay has the potential of jeopardizing the stability of the quasi-coherent states by strongly mixing the different $u$ ladders. As displayed in \cref{fig:hists}(b,d), however, for a limited value of $\gamma$, the intuitive solutions still strongly govern the dynamics. The enhanced mixing of these quasi-coherent states is nevertheless evident in the diminished dwell times on the trajectories, and in that the average intra-cavity photon number (depicted by the white dotted line in \cref{fig:hists}(a) and (b), converges to the detuned empty resonator population in the high-driving limit.

In conclusion, we gained clear insight into the multistability exhibited by the strongly coupled quantum system composed of a driven resonator and a qudit. Beyond the photon blockade regime, where the greatly anharmonic spectrum prevents the system from absorbing excitations, the coupled system can be excited to bright stationary states which can be described as quasi-coherent states with well-defined amplitude and phase that have been determined analytically. The number of the possible stationary states equals the number of the dimensions of the qudit. Identification of them can be a useful resource in sensing applications: the considered small quantum system, when impacted by a weak external signal, can be triggered to undergo a sharp transition between the stationary states with a monitoring output \cite{di2023critical,petrovnin2023microwave,nill2023avalanche}.  

While preparing this manuscript, we became aware of the related work by Karmstrand et al. (arXiv:2403.02417 [2024]), exploring multistability in a similar system in the case of resonant driving, when, in contrast to the offresonant case studied here, the quasi-energies are known exactly.

\acknowledgments
This research was supported by the Ministry of Culture and Innovation and the National Research, Development and Innovation Office within the Quantum Information National Laboratory of Hungary (Grant No. 2022-2.1.1-NL-2022-00004). We are grateful to the HUN-REN Cloud for providing us with suitable computational infrastructure for the simulations.

%\bibstyle{apsrev4-1}
\bibliography{main}

\end{document}